\documentclass[fleqn]{article}
\usepackage{npb}
\usepackage{axodraw}
\usepackage{graphicx}
\usepackage{enumerate}

\newcommand{\be}{\begin{equation}}
\newcommand{\ee}{\end{equation}}
\newcommand{\ba}{\begin{eqnarray}}
\newcommand{\ea}{\end{eqnarray}}

\title{Neutrino Oscillations, SUSY See-Saw Mechanism and Charged Lepton Flavor Violation
\thanks{
Talk presented by R.~R\"uckl at RADCOR 2002,
{\it 6th International Symposium on Radiative Corrections}, 8-13/09/02,
Kloster Banz, Germany}
}
\author{F.~Deppisch$^{\rm a}$, H.~P\"as$^{\rm a}$, A.~Redelbach$^{\rm a}$, 
R.~R\"uckl\address{Institut f\"ur Theoretische Physik und Astrophysik,
Universit\"at W\"urzburg, D-97074 W\"urzburg, Germany}, 
Y.~Shimizu\address{Department of Physics, Nagoya University, Nagoya, 464-8602, Japan}}

\def\lsim{\raise0.3ex\hbox{$\;<$\kern-0.75em\raise-1.1ex\hbox{$\sim\;$}}}
\def\gsim{\raise0.3ex\hbox{$\;>$\kern-0.75em\raise-1.1ex\hbox{$\sim\;$}}} 

\newcommand{\mx}{\left[\begin{array}}
\newcommand{\finmx}{\end{array}\right]} 
\newcommand{\mxp}{\left(\begin{array}} 
\newcommand{\finmxp}{\end{array}\right)} 
\def\beq{\begin{equation}}
\def\eeq{\end{equation}}
\def\bea{\begin{eqnarray}}
\def\eea{\end{eqnarray}}

\def\mathbf#1{\hbox{\bf #1}}
\def\textrm#1{\hbox{#1}}

\def\lsim{\raise0.3ex\hbox{$\;<$\kern-0.75em\raise-1.1ex\hbox{$\sim\;$}}}
\def\gsim{\raise0.3ex\hbox{$\;>$\kern-0.75em\raise-1.1ex\hbox{$\sim\;$}}}
\newcommand {\ignore}[1]{}

\begin{document}

\begin{abstract}
Neutrino oscillations give 
clear evidence for non-vanishing neutrino masses and lepton-flavor 
violation (LFV) in the neutrino sector. This provides strong motivation 
to search for signals of LFV also in the charged lepton sector, 
and to probe the SUSY see-saw mechanism.
We compare the sensitivity of rare 
radiative decays on the right-handed 
Majorana mass scale \(M_R\) with the reach in slepton-pair production
at a future linear collider.
\end{abstract}

\maketitle


\section{Neutrino oscillations}

\begin{figure}[t!]
\setlength{\unitlength}{1cm}
\begin{minipage}[t!]{7.5cm}
\includegraphics[clip,scale=0.45]{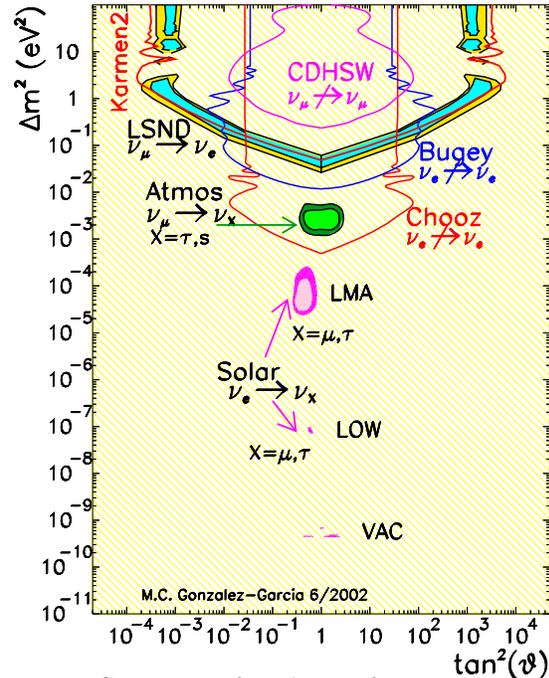}
\vspace*{-2cm}
\caption{Summary of evidences for neutrino oscillations \cite{Gonzalez-Garcia:2002dz}.}
     \label{fig:sum}
\end{minipage}\hfill
\vspace*{-0.7cm}
\end{figure}

In the recent years a rather unique picture of neutrino 
mixing has emerged. As can be seen in 
Fig.~\ref{fig:sum}, large to maximal mixing has been established for
solar and atmospheric neutrinos, while
the third angle is strongly constrained by reactor measurements.
Taking $\theta_{13}=0^o$ and $\theta_{23}=45^o$,
the mixing matrix turns out to be of
the form
\ba{}
V&=&\left(\begin{array}{ccc}
\cos \theta_{12} & \sin \theta_{12} & 0\\
-\frac{\sin \theta_{12}}{\sqrt{2}}
& \frac{\cos \theta_{12}}{\sqrt{2}} & \frac{1}{\sqrt{2}}\\
\frac{\sin \theta_{12}}{\sqrt{2}}
& \frac{-\cos \theta_{12}}{\sqrt{2}} & \frac{1}{\sqrt{2}}\\
\end{array} \right).
\ea
In the following we focus on the highly favored LMA solution with
$\cos \theta_{12}\simeq 0.86$.
In addition, the mass squared differences
\begin{eqnarray}
\Delta m^2_{12} &\simeq& 6 \times 10^{-5} ~~{\rm eV}^2 \\ 
\Delta m^2_{23} &\simeq& 3 \times 10^{-3} ~~{\rm eV}^2 
\end{eqnarray}
have been deduced from the solar and
atmospheric neutrino measurements, respectively. The future
long baseline experiments
KAMLAND and MINOS, as well as experiments at a neutrino factory 
and observations of galactic
supernovae will significantly improve the knowledge of 
these parameters.

\section{SUSY see-saw mechanism and slepton mass matrix}

If three right-handed neutrino singlet fields $\nu_R$
are added to the MSSM particle 
content, one gets additional terms in the superpotential \cite{Casas:2001sr}:
\begin{equation}
W_\nu = -\frac{1}{2}\nu_R^{cT} M \nu_R^c + \nu_R^{cT} Y_\nu L \cdot H_2.
\label{suppot4}
\end{equation}
Here, \(Y_\nu\) is the matrix of neutrino Yukawa couplings, 
$M$ is the right-handed neutrino Majorana mass matrix, and
$L$ and $H_2$ denote the left-handed 
lepton and hypercharge +1/2 Higgs doublets, respectively. 
At energies much below the mass scale of the right-handed neutrinos, 
$W_{\nu}$ leads to the following
mass matrix for the left-handed neutrinos:
\beq\label{eqn:SeeSawFormula}
M_\nu = m_D^T M^{-1} m_D = Y_\nu^T M^{-1} Y_\nu (v \sin\beta )^2.
\eeq
Thus, light neutrino masses are 
naturally obtained if the mass scale \(M_R\) of the
matrix \(M\) is much larger than the scale of the Dirac mass 
matrix \(m_D=Y_\nu \langle H_2^0 \rangle\) with
\(\langle H_2^0 \rangle = v\sin\beta\) being the appropriate Higgs v.e.v., 
\(v=174\)~GeV and 
\(\tan\beta =\frac{\langle H_2^0\rangle}{\langle H_1^0\rangle}\).
The matrix $M_{\nu}$ is diagonalized by the unitary MNS matrix \(U\):
\begin{eqnarray}\label{eqn:NeutrinoDiag}
U^T M_\nu U \!\!\! &=& \!\!\! \textrm{diag}(m_1, m_2, m_3),\\
U \!\!\! &=& \!\!\! \textrm{diag}(e^{i\phi_1}, e^{i\phi_2},1)V(\theta_{12}, 
\theta_{13}, \theta_{23}, \delta), \nonumber
\end{eqnarray}
where \(\phi_1\) and \(\phi_2\) are Majorana phases and $m_i$ the 
light neutrino mass eigenvalues.

The other neutrino mass eigenstates 
are too heavy to be observed directly. However, they
give rise to virtual corrections 
to the slepton mass matrices 
that can be responsible for observable lepton-flavor violating effects.
In particular, the mass matrix of the charged sleptons is given by
\begin{eqnarray}
 m_{\tilde l}^2=\left(
    \begin{array}{cc}
        m_{\tilde l_L}^2    & (m_{\tilde l_{LR}}^{2})^\dagger \\
        m_{\tilde l_{LR}}^2 & m_{\tilde l_R}^2
    \end{array}
      \right)
\end{eqnarray}
with
\begin{eqnarray*}
  (m^2_{\tilde{l}_L})_{ab}     \!\!\!&=&\!\!\! (m_{L}^2)_{ab} 
+ \delta_{ab}\bigg(m_{l_a}^2 
\nonumber \\ 
&+& m_Z^2 
\cos 2\beta \left(-\frac{1}{2}+\sin^2\theta_W \right)\bigg) 
\label{mlcharged} \\
  (m^2_{\tilde{l}_{R}})_{ab}     
\!\!\!&=&\!\!\! (m_{R}^2)_{ab} 
\nonumber \\
\!\!\!&+&\!\!\! \delta_{ab}(m_{l_a}^2 - m_Z^2 \cos 2\beta 
\sin^2\theta_W) \label{mrcharged} \\
 (m^{2}_{\tilde{l}_{LR}})_{ab} 
\!\!\!&=&\!\!\! A_{ab}v\cos\beta-\delta_{ab}m_{l_a}\mu\tan\beta.
\end{eqnarray*}
When $m^2_{\tilde{l}}$
is run from the GUT scale \(M_X\) to the electroweak scale 
one obtains, in mSUGRA,
\begin{eqnarray}
m_{L}^2\!\!\!&=&\!\!\!m_0^2\mathbf{1} + (\delta m_{L}^2)_{\textrm{\tiny MSSM}} + \delta m_{L}^2 \label{left_handed_SSB} \\
m_{R}^2\!\!\!&=&\!\!\!m_0^2\mathbf{1} + (\delta m_{R}^2)_{\textrm{\tiny MSSM}} + \delta m_{R}^2 \label{right_handed_SSB}\\
A\!\!\!&=&\!\!\!A_0 Y_l+\delta A_{\textrm{\tiny MSSM}}+\delta A \label{A_SSB},
\end{eqnarray}
where $m_{0}$ is the common soft SUSY-breaking scalar mass and $A_{0}$ the 
common trilinear coupling. The terms 
\((\delta m_{L,R}^2)_{\textrm{\tiny MSSM}}\) and \(\delta A_{\textrm{\tiny MSSM}}\) are well-known flavor-diagonal corrections.
In addition, the evolution generates off-diagonal terms 
which
in leading-log approximation are given by \cite{Hisano:1999fj}
\begin{eqnarray}\label{eq:rnrges}
  \delta m_{L}^2 \!\!\!&=&\!\!\! -\frac{1}{8 \pi^2}(3m_0^2+A_0^2)(Y_\nu^\dag Y_\nu) 
\ln\left(\frac{M_X}{M_R}\right) \label{left_handed_SSB2}\\
  \delta m_{R}^2\!\!\! &=&\!\!\! 0  \\
  \delta A\!\!\! &=&\!\!\! -\frac{3 A_0}{16\pi^2}(Y_l Y_\nu^\dag Y_\nu) \ln\left(\frac{M_X}{M_R}\right).
\end{eqnarray}

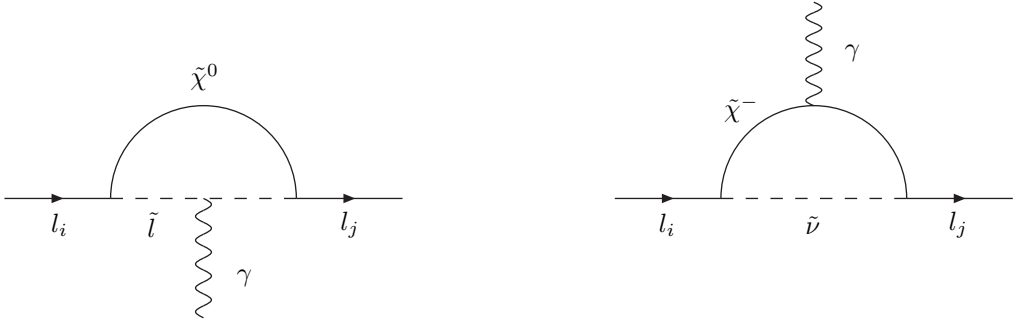
\begin{figure*}[!t]
\begin{center}
\begin{picture}(400,100)(-200,-25)
\ArrowLine(-190,0)(-150,0)
\Text(-170,-10)[]{$l_{i}$}
\DashLine(-150,0)(-80,0){5}
\Text(-135,-10)[]{$\tilde{l}$}
\Photon(-115,0)(-115,-45){3}{5}
\Text(-100,-30)[]{$\gamma$}
\ArrowLine(-80,0)(-40,0)
\Text(-60,-10)[]{$l_{j}$}
\CArc(-115,0)(35,0,180)
\Text(-115,46)[]{$\tilde{\chi}^{0}$}
\ArrowLine(40,0)(80,0)
\DashLine(80,0)(150,0){5}
\ArrowLine(150,0)(190,0)
\Text(60,-10)[]{$l_{i}$}
\Text(115,-10)[]{$\tilde{\nu}$}
\Photon(115,35)(115,74){3}{5}
\Text(130,55)[]{$\gamma$}
\Text(170,-10)[]{$l_{j}$}
\CArc(115,0)(35,-360,-180)
\Text(88,34)[]{$\tilde{\chi}^{-}$}
\end{picture} 
\caption{\label{lfv_lowenergydiagrams} Diagrams for $l_{i}^{-}\rightarrow l_{j}^{-}\gamma$ in the MSSM}
\end{center}
\vspace*{-0.5cm}
\end{figure*}
The product of the neutrino Yukawa couplings
$Y_\nu^\dagger Y_\nu$ entering these corrections can be 
determined \cite{Casas:2001sr} by inverting (\ref{eqn:NeutrinoDiag}),
\begin{eqnarray}\label{eqn:yy}
 Y_\nu \!=\! \frac{\sqrt{M_R}}{v\sin\beta}\!\cdot\!R\!\cdot\!\textrm{diag}(\sqrt{m_1},\sqrt{m_2},\sqrt{m_3})\!\cdot\! U^\dagger,
\end{eqnarray}
using the neutrino data sketched in section 1
as input, and evolving  the result
to the unification scale $M_X$.
In the above, we have chosen a basis in which 
the charged lepton Yukawa couplings are diagonal and have assumed degenerate 
right-handed Majorana masses.
If the unknown complex orthogonal matrix \(R\) in (\ref{eqn:yy})
is real, 
it drops out from the product $Y_\nu^\dagger Y_\nu$. Moreover, the latter is
also independent of \(\phi_1\) and \(\phi_2\). 
In what follows we refer to
this convenient case which suffices for the present discussion.

For hierarchical (a) and degenerate (b) neutrinos one then obtains
\begin{eqnarray}
\mbox{(a) }
 \left(Y_{\nu}^{\dagger}Y_{\nu}\right)_{ab}\!\!\!\!\! &\approx& \!\!\!\!\!
\frac{M_{R}}{v^{2}\sin^{2}\beta}\nonumber \\
  \!\!\!\!\!\!\!\!\!\!\!\!\!\!\!\!\!\!\!\!\!\!\!\!& &\!\!\!\!\!\!\!\!\!\!\!\!\!\!\!\!\!\!\!\!\!\!\!\!\!\!\!\!\!\!\!\!\!\!\!
  \times \left(\sqrt{\Delta m^{2}_{12}} U_{a2}U_{b2}^{*} + \sqrt{\Delta m_{23}^2}U_{a3}U_{b3}^{*}\right) 
\label{llcorrectionhier} \\
\mbox{(b) } \left(Y_{\nu}^{\dagger}Y_{\nu}\right)_{ab}\!\!\!\!\! &\approx&\!\!\!\!\!
\frac{M_{R}}{v^{2}\sin^{2}\beta}
\bigg( m_{1}\delta_{ab}  \nonumber\\
  \!\!\!\!\!\!\!\!\!\!\!\!\!\!\!\!\!\!\!\!\!\!\!\!& &\!\!\!\!\!\!\!\!\!\!\!\!\!\!\!\!\!\!\!\!\!\!\!\!\!\!\!\!\!\!\!\!\!\!\!
+ \frac{\Delta m^{2}_{12}}{2 m_1} U_{a2}U_{b2}^{*} + \frac{\Delta m^{2}_{23}}{2m_1} U_{a3}U_{b3}^{*}\bigg), 
\label{llcorrectiondeg}
\end{eqnarray}
where \(\Delta m_{ij}^2=m^2_j-m^2_i\) and, in (b), 
\(m_1^2\gg\Delta m^2_{23}\gg\Delta m^2_{12}\).
Upon diagonalization, the flavor off-diagonal corrections in 
(\ref{left_handed_SSB})-(\ref{A_SSB}) generate flavor-violating couplings
of the slepton mass eigenstates.
It is interesting to note that in the case of degenerate neutrino 
masses, the lepton-flavor violating effects are suppressed by 
\(\sqrt{\Delta m_{ij}^2}/m_1\) relative to the hierarchical case.

\begin{center}
\begin{figure*}[!t]
\( \quad\quad\sum_{a,b} \) 
\begin{picture}(220,50)(-15,48)
  \ArrowLine(50,50)(10,90)         \Text(10,95)[r]{$e^{+}$}
  \ArrowLine(10,10)(50,50)         \Text(10,5)[r]{$e^{-}$}                          \Vertex(50,50){2}
  \Photon(50,50)(100,50){3}{5}     \Text(75,54)[b]{\(\gamma,Z\)}                    \Vertex(100,50){2}
  \DashArrowLine(130,80)(100,50){5}\Text(113,70)[b]{\(\tilde{l}^{+}_{b}\)}      \Vertex(130,80){2}
  \ArrowLine(160,100)(130,80)      \Text(162,100)[l]{\(l_{j}^{+}\)}
  \ArrowLine(130,80)(160,60)       \Text(162,60)[l]{\(\tilde{\chi}^{0}_{\beta}\)}
  \DashArrowLine(100,50)(130,20){5}\Text(113,30)[t]{\(\tilde{l}_{a}^{-}\)}     \Vertex(130,20){2} 
  \ArrowLine(130,20)(160,40)       \Text(162,40)[l]{\(l^{-}_{i}\)}
  \ArrowLine(160,0)(130,20)        \Text(162,0)[l]{\(\tilde{\chi}_{\alpha}^{0}\)}
\end{picture}\(+\sum_{\gamma,a,b}\)
\begin{picture}(90,50)(-10,48)
  \ArrowLine(20,80)(0,100)         \Text(0,105)[r]{$e^{+}$}
  \ArrowLine(0,0)(20,20)           \Text(0,-5)[r]{$e^{-}$} 
  \ArrowLine(20,20)(20,80)         \Text(22,50)[l]{\(\tilde\chi_\gamma^{0}\)} \Vertex(20,20){2} \Vertex(20,80){2}
  \DashArrowLine(40,100)(20,80){5} \Text(28,93.5)[b]{\(\tilde{l}_{b}^{+}\)}          \Vertex(40,100){2}
  \ArrowLine(70,100)(40,100)       \Text(72,100)[l]{\(l_{j}^{+}\)}
  \ArrowLine(40,100)(70,70)        \Text(70,70)[l]{\(\tilde{\chi}_{\beta}^{0}\)}
  \DashArrowLine(20,20)(40,0){5}   \Text(28,5)[t]{\(\tilde{l}^{-}_{a}\)}             \Vertex(40,0){2}
  \ArrowLine(40,0)(70,30)          \Text(70,0)[l]{\(\tilde{\chi}_{\alpha}^{0}\)}
  \ArrowLine(70,0)(40,0)           \Text(70,30)[l]{\(l^{-}_{i}\)}
\end{picture}
\vspace{1cm}
\caption{Diagrams for $e^+e^-\to \tilde{l}_a^-\tilde{l}^+_b \to 
l_i^-l^+_j \tilde{\chi}^0_\alpha\tilde{\chi}^0_\beta$}\label{e+e-_diags}
\end{figure*}
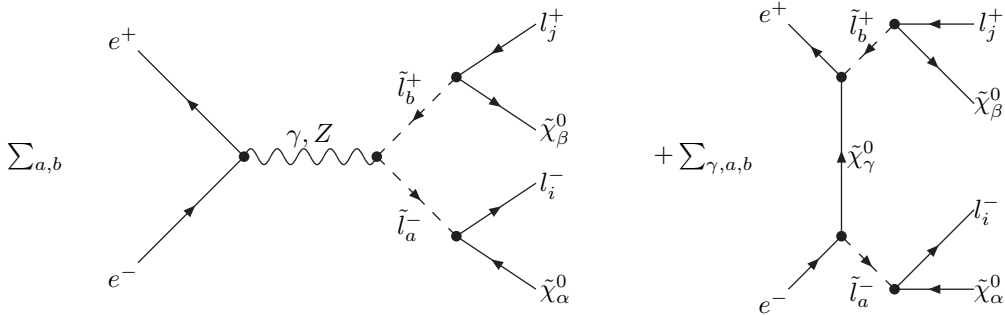
\end{center}

\section{Charged lepton flavor violation}

If only right-handed neutrinos are added to the standard model,
charged lepton-flavor violating processes are suppressed due to the small 
neutrino masses 
\cite{petcov}. However, as outlined above, in supersymmetric models one has 
new sources of lepton-flavor violation, which may give rise 
to observable effects (see also \cite{othertalks}).

At low energies, the flavor off-diagonal correction 
(\ref{left_handed_SSB2}) induces
the radiative decays 
\(l_i\rightarrow l_j \gamma\).
From the photon penguin diagrams shown in 
Fig.~\ref{lfv_lowenergydiagrams} with 
charginos/sneutrinos or neutralinos/charged sleptons in the loop, one 
derives decay rates \cite{Casas:2001sr,Hisano:1999fj}
\begin{equation}
\Gamma(l_i \rightarrow l_j \gamma) 
\propto \alpha^3 m_{l_i}^5 \frac{|(\delta m_L)^2_{ij}|^2}{\tilde{m}^8} 
\tan^2 \beta,
\end{equation}
where $\tilde m$ characterizes sparticle masses.
Because of the dominance of the penguin contributions, the process  
\(\mu\to 3e\), and also \(\mu\)-\(e\) 
conversion in nuclei
is directly related to \(\mu\to e \gamma\), e.g.,
\begin{equation}
\frac{Br(\mu\rightarrow 3e)}{Br(\mu\rightarrow e \gamma)} 
\approx \frac{\alpha}{8\pi}\frac{8}{3}
\left(\ln\frac{m_{\mu}^{2}}{m_{e}^{2}}-\frac{11}{4}\right). \label{mu3erel}
\end{equation}

At high energies, a feasible test of LFV is provided 
by the process
$e^+e^- \to \tilde{l}_a^-\tilde{l}^+_b\to 
l_i^-l^+_j\tilde{\chi}^0_\alpha\tilde{\chi}^0_\beta$.
From the Feynman graphs displayed in Fig.~\ref{e+e-_diags}, one can see that 
lepton-flavor violation can occur in 
production and decay vertices.
The helicity amplitudes \(M_{ab}\) for the pair 
production of \(\tilde{l}_{a}^-\) 
and \(\tilde{l}_b^+\), and the corresponding decay amplitudes 
\(M_a^-, M_b^+\) are given explicitly 
in \cite{inprep}. 
The complete amplitude squared is given by
\begin{eqnarray}
  |M|^2\!\!\!&=&\!\!\!\sum_{abcd}
   (M_{ab} M_{cd}^{*})(M^-_a M_c^{-*})
(M^+_b M_d^{+*})\nonumber\\
  &\times&\!\!\! \frac{\pi}{2(\tilde m_a \Gamma_a + \tilde
m_c \Gamma_c + 
i \Delta \tilde m^2_{ac})} \nonumber\\
  &\times&\!\!\! \frac{\pi}{2(\tilde m_b \Gamma_b + 
\tilde m_d \Gamma_d + i \Delta 
\tilde m^2_{bd})} \nonumber\\
  &\times&
(\delta(p_3^2-\tilde m_a^2)+\delta(p_3^2-\tilde m_c^2)) \nonumber \\
&\times&
(\delta(p_4^2-\tilde m_b^2)+\delta(p_4^2-\tilde m_d^2))
\label{full_M_squared}
,
\end{eqnarray}
where the sum runs over all internal slepton mass eigenstates,
$\Delta \tilde m^2_{ac} = \tilde m^2_{a}-\tilde m^2_{c}$,
and the
narrow width approximation has been used.


\section{Numerical Results}

For our numerical investigations we have
focussed on the mSUGRA benchmark scenarios proposed in 
\cite{Battaglia:2001zp} for linear collider studies.
These models are consistent with direct SUSY searches, Higgs searches,
$b \to s \gamma$, and astrophysical constraints.
Among them, only the benchmark scenarios B, C, G, and 
I have sleptons which are light
enough to be pair-produced 
at \(e^+e^-\) colliders with c.m. energies \(\sqrt{s}=500\div800\)~GeV. 
On the other hand, 
in the case of the rare lepton decays, where the  
SUSY particles only enter virtually, we have investigated all mSUGRA  
scenarios specified in \cite{Battaglia:2001zp}.

For the neutrino input we have taken 
the global three neutrino LMA fit given in \cite{Gonzalez-Garcia:2001sq} 
and have varied $\Delta m^2_{ij}$ and $\theta_{ij}$
within the 90\% CL intervals.
In order to demonstrate possible future improvements, we have also considered 
uncertainty intervals 
as expected from future experiments \cite{Deppisch:2002vz}.
For hierarchical neutrinos the lightest neutrino mass is assumed to lie in
the interval
\(m_1 \approx 0-0.03\)~eV.
For degenerate neutrinos we take 
\(m_1 \approx 0.3^{+0.11}_{-0.16} \)~eV \cite{Deppisch:2002vz}.
The only free parameter is the Majorana mass scale $M_R$.

\begin{figure}[t!]
\setlength{\unitlength}{1cm}
\begin{minipage}[t!]{7.5cm}
\includegraphics[clip,scale=0.48]{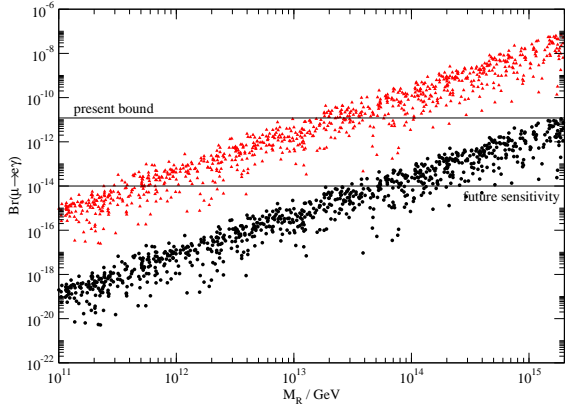}
\vspace*{-1cm}
\caption{Branching ratio of \(\mu \rightarrow e\gamma\) for hierarchical 
neutrino masses  and uncertainties of future neutrino experiments 
in the mSUGRA scenarios L and H leading to the largest and  
smallest signals, respectively.}
     \label{fig:emu_hier_fut}
\end{minipage}\hfill
\end{figure}
\begin{figure}[!t]
\setlength{\unitlength}{1cm}
\begin{minipage}{7.5cm}
\includegraphics[clip,scale=0.48]{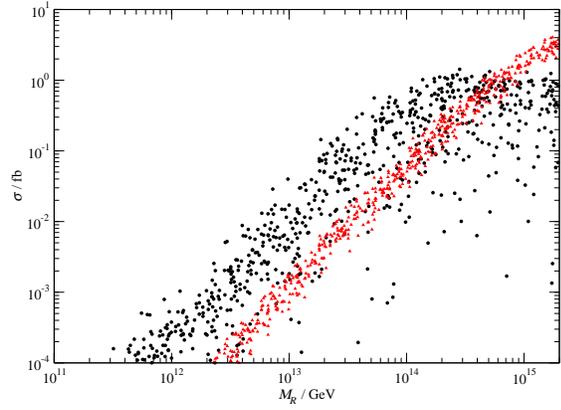}
\vspace*{-0.8cm}
\caption{Cross sections for \(e^+e^- \to e^-\mu^+ +2\tilde\chi_1^0\) 
(circles) and  
\(e^+e^- \to \mu^-\tau^+ +2\tilde\chi_1^0\) (triangles) at
\(\sqrt{s}=500\) GeV
for the mSUGRA scenario B and the same neutrino input as in Fig. 
\ref{fig:emu_hier_fut}.
\label{fig:ep}}
\end{minipage}\hfill
\end{figure}

Fig.~\ref{fig:emu_hier_fut} shows
the dependence of \(Br(\mu\rightarrow e\gamma)\) on \(M_R\) 
for the mSUGRA scenarios L and H, which lead to the largest and smallest 
branching ratios, respectively, in the set of models considered. One sees that
a positive signal between the present bound and the minimum branching 
ratio observable in the new PSI experiment would imply a value of $M_R$ between $5\cdot 10^{11}$~GeV and $5\cdot 10^{15}$~GeV.

In comparison to $\mu \to e \gamma$ the channel
\(\tau\to\mu\gamma\)
is less affected by the neutrino uncertainties. On the other hand,
with a sensitivity in the range \(Br(\tau\to\mu\gamma)=10^{-6}\div 10^{-9}\)
one can only probe $M_R$ above $2\cdot 10^{13}$~GeV
\cite{Deppisch:2002vz}.

\begin{figure}[!t]
\begin{minipage}{7.5cm}
\includegraphics[clip,scale=0.3]{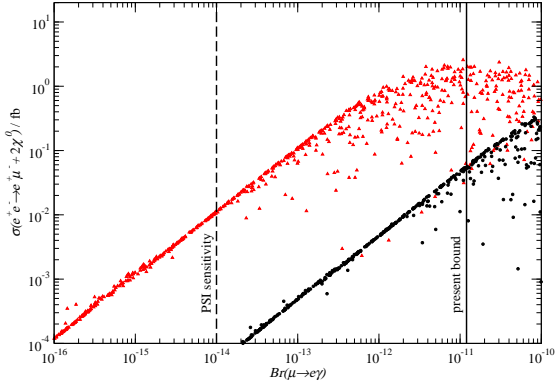}
\caption{Correlation between \(Br(\mu\to e\gamma)\) and   \(\sigma(e^+e^- \to e^-\mu^+ +
2\tilde\chi_1^0)\) at \(\sqrt{s}=800\) GeV for the mSUGRA 
scenarios I (circles) 
and C (triangles) and the same neutrino input as in Fig. 
\ref{fig:emu_hier_fut}. \label{fig:emu_lowhigh}}
\end{minipage}
\end{figure}
In
Fig.~\ref{fig:ep} the cross-sections for \(e^+e^-\rightarrow e^-\mu^++2\tilde{\chi}^0_1\) and 
\(e^+e^-\rightarrow \mu^-\tau^++2\tilde{\chi}^0_1\)  
are plotted versus \(M_R\) at \(\sqrt{s}=500\)~GeV. 
The channel \(e^-\tau^++2\tilde{\chi}_1^0\) is strongly suppressed by the small mixing angle 
\(\theta_{13}\), and therefore more difficult to observe.

The standard model background mainly comes from $W$ production in
$e^+e^-\rightarrow l^-_i l^+_j \bar{\nu}_i \nu_j$. With a beam-pipe cut 
of 10 degrees but no other cuts the cross 
section for the $e\mu$ final state is roughly
$100$~fb at $\sqrt{s}=500$~GeV. 
The MSSM background is dominated by
chargino/slepton production in 
$e^+e^-\rightarrow l^-_il^+_j  +2\tilde{\chi}^0_1 +2(4)\nu$ with
a total cross section of
$0.01\div 0.02$~fb at $\sqrt{s}=500$~GeV for
$l^-_il^+_j=e^-\mu^+$.
This background should be manageable, as the sparticle properties 
will be known 
by the time such an experiment can be performed.

As a final result we point out the
very interesting and useful  
correlation between LFV in the radiative decays and in  high-energy 
slepton-pair production. This is illustrated in Fig.~\ref{fig:emu_lowhigh} 
for \(Br(\mu\to e \gamma)\) and 
\(e^+e^-\rightarrow e^-\mu^++2\tilde{\chi}_1^0\).
In this comparison the neutrino uncertainties 
almost drop out, 
while the sensitivity to the mSUGRA parameters remains.

\section{Conclusions}
The SUSY see-saw mechanism can be tested by lepton-flavor violating 
processes involving charged leptons.
Here we have concentrated on the sensitivity of 
$Br(l_{i}\rightarrow l_{j}\gamma)$ and 
\(\sigma(e^+e^-\rightarrow \tilde{l}_a^-\tilde{l}^+_b\rightarrow 
l^-_i l^+_j + /\!\!\!\!E)\) 
on the Majorana mass scale $M_{R}$. 
Assuming the LMA solution for solar neutrino 
oscillations we have illustrated the impact of the uncertainties in the 
neutrino parameters. 
Furthermore, using post-LEP mSUGRA scenarios we have investigated the
strong dependence of LFV signals
on the mSUGRA parameters.
In the case of hierarchical neutrino masses, a measurement of 
$Br(\mu\rightarrow e\gamma)\approx 
10^{-14}$ would probe $M_{R}$ in the range $3 \cdot 10^{11}\div
4 \cdot 10^{14}$~GeV, 
depending on the mSUGRA scenario. 
For degenerate neutrinos a similar signal would require 
$M_R > 2\cdot 10^{12}$~GeV. 
However, if the mSUGRA 
scenario is known, 
one may actually be able to determine \(M_R\) within a factor of 
10. In comparison, 
a measurement of $Br(\tau\rightarrow \mu\gamma)
\approx 10^{-9}$ would imply \(M_R\) larger than $2\cdot 10^{13}$~GeV
and, for a given scenario, allow to determine  \(M_R\)
within a factor of 
2. 
Finally, we find that \(Br(\mu\rightarrow e \gamma)=10^{-14} \div 10^{-11}\) 
implies  
\(\sigma(e^+e^-\to e^-\mu^++2\tilde{\chi}_1^0)=10^{-2}\div 1\)~fb 
at $\sqrt{s}=800$~GeV in the most favorable mSUGRA scenario C.

\section*{Acknowledgements}   
This work was supported by the Bundesministerium f\"ur Bildung und 
Forschung (BMBF, Bonn, Germany) under 
the contract number 05HT1WWA2.

\end{document}